\documentclass[oldversion,letter]{aa}

\usepackage{graphicx}

\usepackage{txfonts}

\usepackage{color}

\usepackage{natbib}

\usepackage{color}

\def\sss{\scriptscriptstyle}

\def\L{{\sss \!L}}

\def\ISCO{\mathrm{ISCO}}
\def\NS{\mathrm{NS}}

\definecolor{gray}{rgb}{.6,.6,.6}
\definecolor{green}{rgb}{0,.6,0}
\definecolor{red}{rgb}{.9,0,0}

\begin{document}

\title{ {
Appearance of innermost stable circular orbits of accretion discs around rotating neutron stars
}}

\author
{Gabriel T\"or\"ok, Martin Urbanec, Karel Ad\'amek \& Gabriela Urbancov\'a}

\institute{Institute of Physics, Faculty of Philosophy and Science, Silesian
  University in Opava, Bezru\v{c}ovo n\'{a}m. 13, CZ-74601 Opava, Czech Republic
 }
  
\date{Received / Accepted}

\authorrunning{G. T\"or\"ok et al.}
\titlerunning{On neutron stars exhibiting ISCO}
 
\date{Received: 12 February 2014 / Accepted: 10 March 2014}

\abstract
{The innermost stable cicular orbit (ISCO) of an accretion disc orbiting a neutron star (NS) is often assumed a unique prediction of general relativity. However, it has been argued that ISCO also appears around highly elliptic bodies described by Newtonian theory. In this sense, the behaviour of an ISCO around a rotating oblate neutron star is formed by the interplay between relativistic and Newtonian effects. Here we briefly explore the consequences of this interplay using a straightforward analytic approach as well as numerical models that involve modern NS equations of state. We examine the ratio $K$ between the ISCO radius and the radius of the neutron star. We find that, with growing NS spin, the ratio $K$ first decreases, but then starts to increase. This non-monotonic behaviour of $K$ can give rise to a neutron star spin interval in which ISCO appears for two very different ranges of NS mass. This may strongly affect the distribution of neutron stars that have an ISCO 
 (ISCO-NS). When (all) neutron stars are distributed around a high mass $M_0$, the ISCO-NS spin distribution is roughly the same as the spin distribution corresponding to all neutron stars. In contrast, if $M_0$ is low, the ISCO-NS distribution can only have a peak around a high value of spin. Finally, an intermediate value of $M_0$ can imply an ISCO-NS distribution divided into two distinct groups of slow and fast rotators. Our findings have immediate astrophysical applications. They can be used for example to distinguish between different models of high-frequency quasiperiodic oscillations observed in low-mass NS X-ray binaries.}


\maketitle
 
\section{Introduction}
\label{section:introduction}

In Newtonian theory, circular trajectories of test particles orbiting around a spherical central body of mass $M$ are stable to small radial perturbations at any external radii $r$. The same trajectories calculated using a general relativistic description exhibit an instability below the critical radius of the marginally stable cicular orbit $r_{\mathrm{ms}}$ \citep[e.g.,][]{bar-etal:1972}. This radius is frequently considered as the innermost stable circular orbit (ISCO) of an accretion disc that orbits a black hole or a neutron star (NS),  $r_{\mathrm{ISCO}}= r_{\mathrm{ms}}$. The ISCO is often assumed as a unique prediction of Einstein's general relativity. Several decades after the ISCO concept has been introduced, it was argued that ISCO also appears around highly elliptic bodies described by the Newtonian theory \citep[see works of][]{zdu-gou:2001,ams-etal:2002,klu-ros:2013}. In this sense, several phenomena related to rotating oblate neutron stars can be well understood in terms of the interplay between the effects of general relativity and Newtonian theory. Namely, as in the Lense-Thirring or Kerr spacetime, the position of ISCO decreases with growing angular momentum $j$ of the neutron star (given by the NS spin). At the same time, it increases with the increasing influence of the NS quadrupole moment $q$ \citep[which determines the oblateness parameter $\tilde{q}\equiv q/j^2$, see, e.g.,][]{urb-etal:2013}. 

In this paper we explore the consequences of this interplay on the behaviour of NS compactness. Finally, we focus on low-mass X-ray binary systems (LMXBs) and find astrophysical implications for the distribution of neutron stars with an ISCO (ISCO-NS).

\begin{figure*}[t]
\begin{center}
\includegraphics[width=1\hsize]{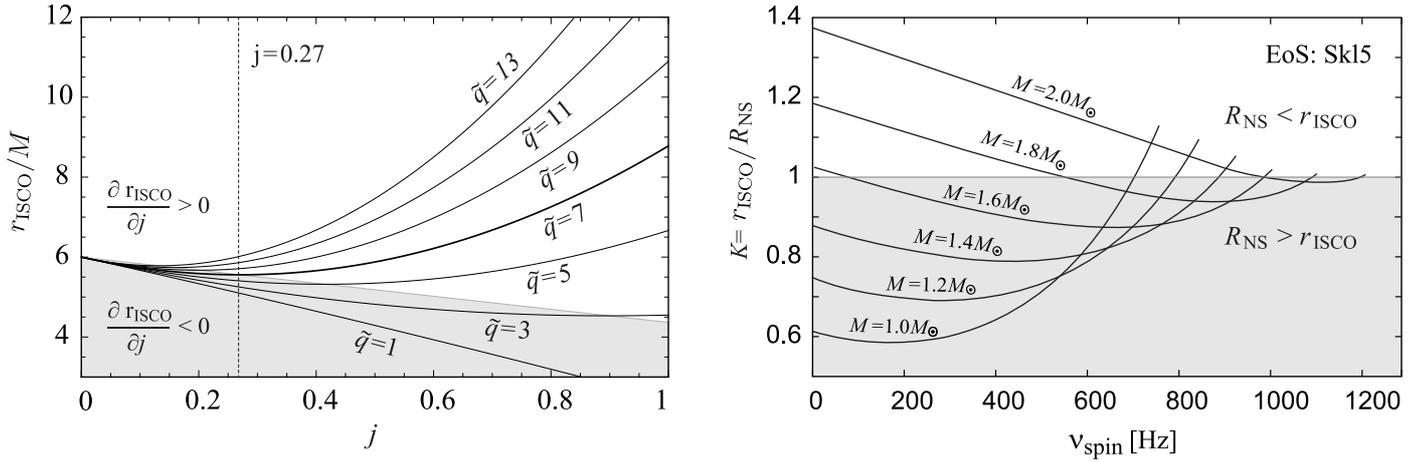}
\end{center}
\caption{\small{Left: Dependence of the geodesic ISCO radius $r_{\mathrm{ISCO}}= r_{\mathrm{ms}}$ on the NS angular momentum $j$ and the oblateness factor $\tilde{q}$. The shaded area indicates the Kerr-like region where $r_{\mathrm{ISCO}}$ decreases with increasing $j$. We note that for the concrete NS models we assume here, a fixed value of $\tilde{q}$ corresponds to a fixed NS central density $\rho$. Moreover, in the limit of $\tilde{q}=1$ the NS spacetime approaches the Kerr spacetime and values of $\tilde{q}\in(1,\,3)$ correspond to the highest possible values of $\rho$ (the oblateness parameter $\tilde{q}$ is also called the Kerr parameter). Right: Dependence of compactness $K$ on the NS spin frequency $\nu_{\mathrm{spin}}$ and mass $M$. The shaded area indicates the region of $K<1$ where the ISCO does not appear above the NS surface.}} 
\label{figure:risco}
\end{figure*}

\section{Behaviour of ISCO radius and NS compactness}
\label{section:radius}

Numerous works have investigated the behaviour of test particle motion, Keplerian frequencies, and the properties of ISCO in the vinicity of a rotating neutron star \citep[e.g.][]{klu-etal:1990,shi-sas:1998,pac-etal:2006,pac-etal:2012,bak-etal:2012,gut-etal:2013}. It has been shown that for geodesic motion, these are well described by formulae derived considering NS spacetime approximated by the Hartle-Thorne metric \citep[][]{har-tho:1968,abr-etal:2003,ber-etal:2005}. Here we explore in more detail the behaviour of function $r_{\mathrm{ISCO}}(j\,,~q)$ calculated by using this approximation.\footnote{The appropriate explicit formulae determining the function $r_{\mathrm{ISCO}}$ are too long to be included in this Letter. They are given, e.g., in \cite{abr-etal:2003} or \cite{tor-etal:2008}.} We find that for $\tilde{q}\gtrsim2.8$, $r_{\mathrm{ISCO}}$ has a minimum when the angular momentum and the quadrupole moment of the central object are related as follows:
\begin{equation}
\tilde{q}_{\mathrm{min}}= 1 + \frac{32\left(9\sqrt{6} + 4j_{\mathrm{min}}\right)}{135 j_{\mathrm{min}} \left[4608 \log(3/2)-1865\right]}\,.
\end{equation}
The smallest allowed ISCO radius is then given by
\begin{equation}
r_{\mathrm{min}}= 6 - 2j_{\mathrm{min}}\sqrt{2/3}\,.
\end{equation}
In the left panel of Figure~\ref{figure:risco} we illustrate the behaviour of ISCO and the turning points $[r_\mathrm{min},j_\mathrm{min} ]$ for several values of $\tilde{q}$. Apparently, for $\tilde{q}\gtrsim7$, the turning points arise for low angular momentum values, $j\lesssim0.27$. 

\subsection{ISCO calculated for particular NS models}

So far we have only considered the behaviour of ISCO without focusing on its relation to NS models. Models of neutron stars based on modern equations of state (EoS) have been extensively developed through the use of various numerical methods and codes \citep[see][for a review]{lat-pra:2001,lat-pra:2007}. Results of the published studies indicate that in contrast to the behaviour of $r_{\mathrm{ISCO}}$, the NS radius $R_{\NS}$ very slowly evolves with the NS spin. Taking this into account, one can expect that the non-monotonicity of the ISCO behaviour discussed above may imply a non-monotonicity of the quantity $K\equiv r_{\ISCO}/ R_{\NS}$ (hereafter denoted as compactness factor). Considering concrete NS models we confirm this expectation using the numerical code of \cite{urb-etal:2013}.

In the right panel of Figure~\ref{figure:risco} we plot the NS compactness factor $K$ calculated for the SKl5 Skyrme EoS \citep[e.g.,][]{rik-etal:2003}. We assume in the figure several values of the neutron star mass $M$ and spin frequency $\nu_{\mathrm{spin}}$, which range from the non-rotating limit to highest frequencies corresponding to the mass-shedding limit. To obtain smooth curves we have computed {$600$} NS configurations. One can clearly see that $K$ is not a monotonic function of $\nu_{\mathrm{spin}}$. Its extrema arise for very different values of $\nu_{\mathrm{spin}}$. For a high neutron star mass, ($M\sim2M_{\sun}$), they arise close to the mass-shedding limit, above 1000Hz. For a low neutron star mass, ($M\sim1M_{\sun}$), they arise for low values of $\nu_{\mathrm{spin}}$, below $\nu_{\mathrm{spin}}=200$Hz. 


\section{Appearance and disappearance of ISCO and the consequences}

Using the right panel of Figure~\ref{figure:risco} we can directly compare the NS radii and the ISCO radii. Apparently, when $M$ is high, the ISCO is located above the NS surface for  almost any $\nu_{\mathrm{spin}}$. On the other hand, when $M$ is low, the ISCO arises for only high values of $\nu_{\mathrm{spin}}$. For an intermediate mass, $M\sim1.6M_{\odot}$, the ISCO only occurs for two short intervals of $\nu_{\mathrm{spin}}$ - one close to the non-rotating limit and the other one at high frequencies.

\begin{figure*}[t]
\begin{center}
\includegraphics[width=1\hsize]{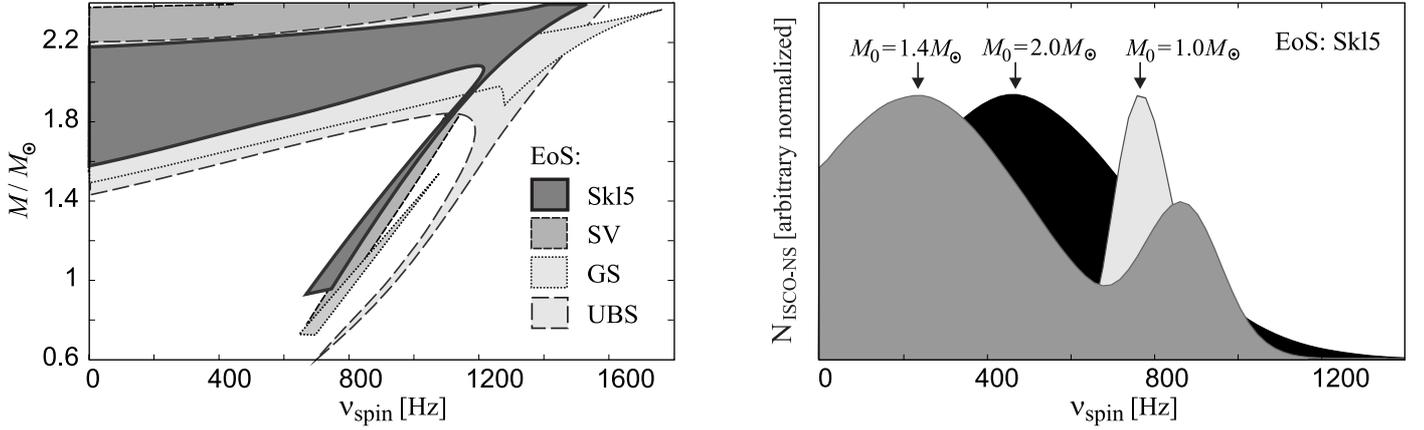}
\end{center}
\caption{\small{Left: Mass-spin regions that allow the appearance of ISCO calculated for SKl5 and three other EoS. The displayed drawing is the result of {$\sim10^5$} computations of NS configurations. Right: ISCO-NS spin distribution expected for the SKl5 EoS and three different referential values of $M_0$.}}
\label{figure:distribution}
\end{figure*}

In the left panel of Figure~\ref{figure:distribution} we show the map of mass--spin regions where the ISCO arises above the NS surface for Skl5 EoS. This map, which relates to curves drawn in the right panel of Figure~\ref{figure:risco}, is emphasized by the thick line and dark colour. Remarkably, for $\nu_{\mathrm{spin}}\sim800$Hz, there are two intervals of $M$ that allow an ISCO appearance. On the other hand, for $\nu_{\mathrm{spin}}\lesssim700$Hz (or $\nu_{\mathrm{spin}}\gtrsim1200$Hz), there is only one such interval of $M$. For the sake of comparison we also consider three other EoS within the same figure: the SV and GS EoS, which represent two more parametrizations of the Skyrme potential \citep[see][and references therein]{rik-etal:2003}, and the UBS EoS \citep[][]{urb-etal:2010:aca}. Apparently, these EoS display a similar behaviour as the Skl5 EoS.

\subsection{Possible double-peaked spin distribution of ISCO-NS}

While it is often speculated that most neutron stars in LMXBs may have ISCO, little clear observational evidence for this claim has been achieved yet, and the mass and spin distribution of these possible ISCO-NS remains puzzling \citep[see, however][]{bar-etal:2005,bar-etal:2006,tor-etal:2010,tor-etal:2012,wan-etal:2013}. The mass distribution of neutron stars in LMXBs has hardly been explored so far. The mean mass is currently estimated as $M_0\in(1.4M_\odot-1.8M_\odot)$. More accurate estimates on NS mass distribution are available in double NS binary systems where masses are concentrated in a very narrow peak around 1.4$M_\odot$ \citep[see][]{Lat:2012:}.\footnote{This rather brief description of situation does not include recent observations of the pulsar J0348+0432 with mass $2.01 \pm 0.04 M_\odot$, which currently represents the most massive confirmed neutron star \citep{Ant-etal:2013:}.} 
Accurate measurements for NS spin frequencies are available in radio pulsars where the distribution of rotational frequencies is narrowed around two peaks - one is of order of ~Hz, while the other is $\sim$100Hz \citep{Man-etal:2005:}. Spin measurements are much less frequent in LMXBs. {The spin frequencies inferred from the observed X-ray oscillations seem to be significantly higher than those of radio  pulsars, reaching typical values of few hundreds of Hz. Nevertheless, several slowly rotating sources have been discovered as well \citep[e.g., IGR J17480-2446 in globular cluster Terzan X-5 with $\nu_{\mathrm{spin}}=11$Hz,][]{bor-etal:2010,mar-str:2010}. An overview of the currently measured spins of more than 20 LMXB NSs along with detailed references can be found in \cite{lam-bou:2008} and \cite{pat-wat:2012}.} 

Motivated by a simple analytical analysis and justified by the consideration of NS models assuming modern EoS, our results imply strong restrictions on distribution of the ISCO-NS parameters. This distribution could be very different from the (so far unknown) distribution of all NS in LMXBs. We illustrate these restrictions using an example calculated for the Skl5 EoS and the simplified assumption that all NS follow a two dimensional Gaussian distribution in the mass--spin space. For the distribution of all NS we set mean values $M_0\in\left\{1M_{\sun},~1.4M_{\sun},~2M_{\sun}\right\},$ and $\nu_{\mathrm{spin}}^0=500\mathrm{Hz}$ together with scales $\sigma_M=0.2M_{\sun}$ and $\sigma_{\nu\mathrm{spin}}=300$Hz. The subsets of ISCO-NS were then determined by the map depicted in the left panel of Figure~\ref{figure:distribution}. We show the spin distribution of these subsets in the right panel of Figure~\ref{figure:distribution}. We can see that when the neutron stars were distributed around high mass $M_0=2M_{\sun}$, the ISCO-NS spin distribution is approximately the same as the spin distribution of all NS that peak around $\nu_{\mathrm{spin}}^0$. On the other hand, when this mass was low, $M_0=1M_{\sun}$, the ISCO-NS distribution had a peak only around a high value of $\nu_{\mathrm{spin}}$. A different situation occurs for an intermediate value of $M_0=1.4M_{\sun}$, which implies an ISCO-NS distribution divided into two groups of slow and fast rotators.

\section{Discussion and conclusions}
\label{section:conclusions}

Our analysis clearly revealed the non-monotonicity of the dependence of the compactness factor $K$ on NS spin. The occurence of NS spin interval in which the ISCO appears for two very different ranges of NS mass (and the inferred double-peaked ISCO-NS spin distribution) depends on the detailed behaviour of $K$.

For the four EoS discussed here, this behaviour is rather similar. However, normalization of $K$ and its exact dependence on $\nu_{\mathrm{spin}}$ and $M$ depend on the particular EoS. More detailed investigation that takes into account a large set of EoS needs to be performed to make better astrophysical assessments. Moreover, we assumed here the simplified identity $r_{\mathrm{ISCO}}= r_{\mathrm{ms}}$, but various effects might lead to a decrease or increase of $K$ (e.g. viscosity, pressure forces, or magnetic field influence), and the chosen spacetime description can play some role as well \citep[e.g.][]{alp-psa:2008,str-sra:2009,bak-etal:2010,bak-etal:2012,kot-etal:2008}. Despite these uncertainties, which need to be adressed in the subsequent analysis, our findings can be useful in several astrophysical applications. {Clearly, the predictions on the NS parameters would be strong if the ISCO presence or absence in the individual systems were observationally confirmed. Moreover, one can speculate on some consequences for evolution of the spinning-up sources.} 

{Although more development is required, some implications can be consired immediately. One possibility is to use our results in a study aimed to distinguish between different models of high-frequency quasiperiodic oscillations (HF QPOs). There is no consensus as yet on the HF QPO origin, and various models have been proposed \citep[e.g.,][and several others]{alp-sha:1985,lam-etal:1985,mil-etal:1998,psa-etal:1999,ste-vie:1999,abr-klu:2001,tit-ken:2002,rez-etal:2003,pet:2005,zha:2005,sra-etal:2007,stu-etal:2008}. Among numerous ideas,} it was suggested that the strong luminosity variations observed on the kHz time-scales in the neutron star X-ray binaries result from the modulation of accretion flow that heats the boundary layer between the flow and the NS surface \citep[][]{pac:1987,hor:2005,abr-etal:2007}. Motivated by this suggestion, \cite{urb-etal:2010} assumed the Paczynski modulation mechanism for the specific epicyclic resonance QPO model and investigated the implied restrictions on the NS mass and angular momentum. Taking into account our present findings and the general idea that (any) disc oscillations are responsible for exciting the Paczynski modulation mechanism, one might expect that the spin distribution of the HF QPO sources is double-peaked when NS masses are distributed around the intermediate value $M_0\sim1.4M_{\sun}$. 

\section*{Acknowledgments}
We would like to acknowledge the support of the Czech grant GA\v{C}R 209/12/P740 and the project CZ.1.07/2.3.00/20.0071 "Synergy", aimed to foster international collaboration of the Institute of Physics of SU Opava. Furthermore, we acknowledge financial support from the internal grants of SU Opava SGS/11/2013 and SGS/23/2013. We thank the anonymous referee for his or her comments and suggestions that helped to improve the paper.



\end{document}